\documentclass[11pt]{article}

\usepackage{amsmath}
\usepackage{amssymb}

\usepackage{graphicx}
\usepackage[tight,footnotesize]{subfigure}
\usepackage{cite}
\usepackage{url}

\usepackage{color} 

\usepackage[labelfont=bf,labelsep=period,justification=raggedright]{caption}

\makeatletter
\renewcommand{\@biblabel}[1]{\quad#1.}
\makeatother

\date{}

\begin{document}

% Title must be 150 characters or less
\begin{flushleft}
{\LARGE \textbf{A measure of total research impact independent of time and discipline}
\vspace*{0.2em}}
% Insert Author names, affiliations and corresponding author email.
\\
Alberto Pepe$^{1,\ast}$,
Michael J. Kurtz$^{2}$
\\
{\bf 1} Center for Astrophysics, Harvard University, Cambridge, MA
\\
{\bf 2} Center for Astrophysics, Smithsonian Astrophysical
Observatory, Cambridge, MA
\\
%\bf{3} Author3 Dept/Program/Center, Institution Name, City, State, Country
{\bf $\ast$} Corresponding author: apepe@cfa.harvard.edu
\end{flushleft}

% Please keep the abstract between 250 and 300 words
{\small \textbf{Abstract}. Authorship and citation
practices evolve with time and differ by academic discipline. As such,
indicators of research productivity based on citation records are
naturally subject to historical and disciplinary effects. We observe
these effects on a corpus of astronomer career data constructed from a
database of refereed publications. We employ a simple mechanism to
measure research output using author and reference counts available in
bibliographic databases to develop a citation-based indicator of
research productivity. The total research impact (tori) quantifies,
for an individual, the total amount of scholarly work that others have
devoted to his/her work, measured in the volume of research papers. A derived measure, the research
impact quotient (riq), is an age independent measure of an individual's
research ability. We demonstrate that these measures are substantially less
vulnerable to temporal debasement and cross-disciplinary bias than the
most popular current measures. The proposed measures of research
impact, tori and riq, have been implemented in
the Smithsonian/NASA Astrophysics Data System.}

\section*{Introduction}
Measuring the research performance of scholars plays a critical role
in the allocation of scholarly resources at all levels
\cite{bernal,merton,price1963,Lehmann:2006fk,nsf,kurtz05a,eindex}.  A principal
``quantitative'' means of measurement has long been through the use of
citations \cite{Garfield55,desollaprice}.  Citations are routinely
used to evaluate the research productivity of individuals
\cite{hindex,gindex}, journals
\cite{garfield,pinskin76,west,moed,zitt,zitt2}, universities
\cite{leydes,nsf}, and nations \cite{may,king,bonitz}.  The use of
citations to measure research performance involves several confounding
factors which tend to become
more important as the degree of aggregation decreases.  For the
evaluation of individuals, important challenges are:

\begin{description}
\item[Discipline]  Citation practices vary widely among various fields.
Citation rates can vary between disciplines by an order of
magnitude \cite{leydes}; among sub-disciplines in the same discipline they can
vary by a factor of two (see Figure \ref{fig:mVSdisc}, discussed later).

\item[Co-Authorship] A paper can have an arbitrary number of authors, from
one to several thousand.  Should an author of a single authored paper
receive the same credit for a citation as someone who has co-authors?

\item[Age] The number of citations accrued by an individual scales
with the square of his/her career length \cite{hindex,kurtz05b};
thus, a person with a career length of 10 years will have half the
citations of an equal person with a career length of 14.14 years. This
age effect problem is exacerbated by the fact that the two
aforementioned challenges are time dependent. For example, in the
field of astrophysics, both the mean number of references and the mean
number of authors have approximately doubled in the last 20 years.
\cite{Henneken2011,Schulman1997}
\end{description}

Some of the lesser challenges associated with using citations to measure
research productivity of individuals are:
\begin{description}
\item[Self-Citation] If an author cites papers by him/herself should they
count as much as citations from papers by others?

\item[Curation] In addition to having a database of articles and
citations, one must clean and curate its data. For example, 
an analysis of an individual's productivity requires that
one be able to exactly identify the articles written by that
individual. Name changes (e.g., due to marriage) and homonyms (name clashes, where different people have the same name) can make this a serious
problem.

\item[Shot Noise] Sometimes an individual can, almost entirely by
chance, become an author of one or more very highly cited papers,
perhaps as a student.  The citation distribution is a Zipf like power
law, whereby some articles are cited thousands of times more than the
median; clearly, there can be circumstances where a direct count of
citations is not a fair representation of impact.
\end{description}

In a highly influential paper, Hirsch \cite{hindex} proposed a pair of
citation-based measures (\textit{h, m}) which: solve the shot-noise
problem, substantially improve the age problem, and help with the
curation difficulty, discussed above. The Hirsch index, \textit{h}, is
the position in a citation ranked list where the rank equals the
number of citations; absent shot noise {\it h} is obviously proportional
to the square root of the total number of citations, which grows
linearly with career length \cite{hindex,kurtz05b}.  The \textit{m}
quotient is \textit{h} divided by career length, and is a constant
throughout the career of an individual with constant productivity in a
constant environment.

The \textit{h}-index is by far the most widely used
indicator of personal scientific productivity. As such, it has been greatly
reviewed and criticized in specialized literature and innumerable alternatives have been
proposed \cite[for a review]{egghes}. Some notable substitutes of the h-index include: 
the mean number of citations per paper \cite{Lehmann:2006fk}, the \textit{e}-index which complements the h-index for excess citations \cite{eindex}, the \textit{g}-index, similar to {\it h}, but differs for it accounts for the averaged citation count an author has accrued \cite{gindex}, and the highly cited publications indicator \cite{inconsistency}.
Two normalizations of the {\it h}-index which have been proposed in the literature with promising results are by the number of article co-authors \cite{batista}, and by the average number of citations per article per
discipline \cite{radicchi}. The measures proposed in this article use both of these normalizations, combined.

%Yet, the \textit{h}-index remains the most widely available yardstick
%in the hands of scholars to measure their performance. 
%Its strength lies in its simplicity and ease of calculation: a
%scholar ought to know only the citation rate of her most cited papers
%to calculate her \textit{h}-index.

While the \textit{h}-index is a valuable, simple, and effective
indicator of scholarly performance, we find that it is inadequate for
cross-disciplinary and historical comparisons of
individuals. Comparing two scholars from different disciplines or from
different time periods, or with differing co-authorship practices,
based on their \textit{h}-index would very likely yield erroneous
results, simply because citation and authorship practices have changed
(and constantly change) across disciplines and through time.

\section*{Methods}
To investigate the historical and disciplinary effects of the \textit{h}-index, we calculate
individual researcher performance on a virtually complete
astronomy database of 814,505 refereed publications extracted from the
Smithsonian/NASA Astrophysics Data System
(\url{http://adsabs.harvard.edu/}) \cite{kurtz05a}. 
We focus on the careers of 11,036 astronomers with non ambiguous names, with a
publication record of over 20 refereed articles and a career span of
over 10 years, who are either currently active or have a career length
of at least 30 years which started on or after 1950. We define the
beginning of the career as the year of publication of an astronomer's
first refereed article. To begin, we compute the \textit{m}-quotient
on this cohort of astronomers and demonstrate that it is not 
constant over time and across sub-disciplines of astronomy.
Then, we propose a novel measure of research performance, the research
impact quotient (\textit{riq}). We compute \textit{riq} on the
same bibliographic corpus showing that this derived measure eliminates most historical
and disciplinary bias.

\section*{Results}
\subsection*{Temporal debasement and cross-disciplinary bias of current
measures} 
\begin{figure*}[hp!]
\begin{center}
\centerline{\includegraphics[width=1\textwidth]{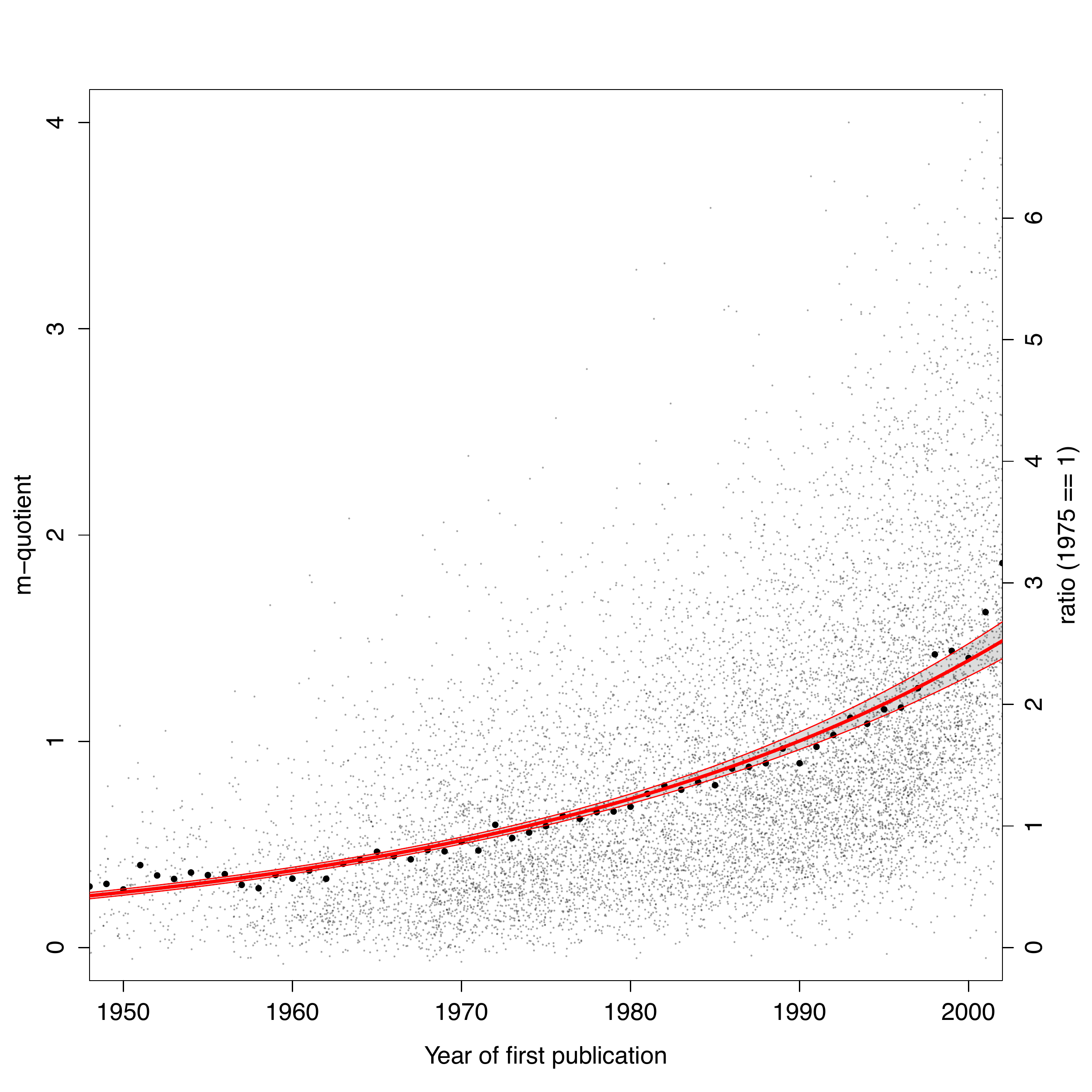}}
\caption{\label{fig:mVSy1} Distribution of astronomers' \textit{m}-quotients as function
  of beginning of career (defined as the year of first refereed publication).}
\end{center}
\end{figure*}

In Figure \ref{fig:mVSy1}, we illustrate the temporal debasement of
the \textit{m}-quotient, defined as $h/y$, where $y$ is the number of years since a scholar's
first publication. Astronomers who began their career in the
1950's have systematically lower \textit{m}s than those who started
their career later on. The red line in Figure \ref{fig:mVSy1} is an
exponential best-fit regression line with slope $b=0.0314$ and a
$0.95$ confidence interval band. Year means are plotted as filled
black circles. In 50 years, the average \textit{m}-quotient has
increased from $0.28$ to $1.62$, with an increase rate of $3.1$\% per
year, and well above the global mean of $x = 0.855$.  (We also run an
identical regression analysis on a cohort of 697 astronomers for whom
we have access to both publication record \textit{and}
Ph.D. dissertation. Using the doctoral graduation year as the starting
point of their career we find similar effects of temporal debasement
--- best-fit regression line has slope $b=0.027$.)

\begin{figure*}[hp!]
\begin{center}
\includegraphics[width=1\textwidth]{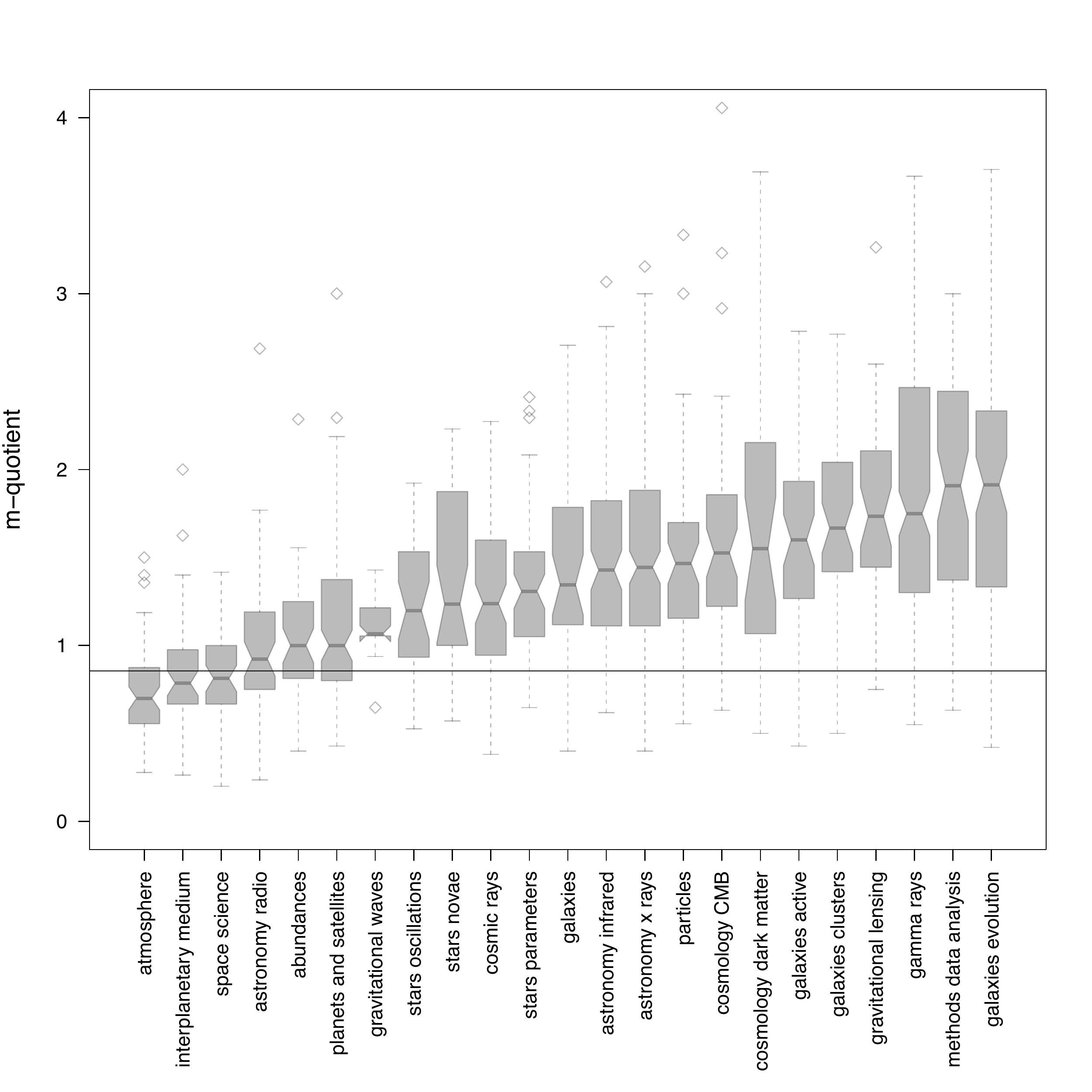}
\caption{\label{fig:mVSdisc} Distribution of astronomers'
  \textit{m}-quotients as function of field of specialization. Each field includes between 30 and 150 astronomers who began their career in the 1990s. The box-and-whisker plot of each field depicts the median (middle notch), lower and upper quartiles (lower and upper hinges), minimum and maximum values (lower and upper whiskers), and outliers.}
\end{center}
\end{figure*}

In Figure \ref{fig:mVSdisc} we show cross-disciplinary bias of the
\textit{m}-quotient.  In the figure, the {\it m}-quotients of
astronomers working in different fields of specialization is displayed as a box-and-whisker
plot. Astronomers' fields of specialization are computed by
simply selecting the single most recurrent keyword used by authors in
their published articles. In order to isolate disciplinary effects, we
only analyze a subset of the corpus which includes $1601$ astronomers
who started their career in the 1990s and who publish in popular
sub-disciplines in this time window (fields with 30 authors or less
are excluded from this analysis). 
The dashed line in Figure \ref{fig:mVSdisc} shows the global
mean \textit{m}-quotient for all authors in the corpus
($x=0.855$). 
We find that for only a small
portion of sub-disciplines (6 out of 23) does the global mean {\it
m}-quotient fall within the discipline-specific upper or lower
quartiles (``atmosphere'' through ``planets and
satellites''). Astronomers who publish in all the other fields have
systematically higher \textit{m}-quotients than the global average, as evinced by higher median \textit{m}-quotients for fields ``gravitational waves'' to ``galaxies evolution''.

Differences so large across time and disciplines make comparison of
individuals, such as in promotion and tenure decisions, quite
difficult.  Over time, a 3.1\% yearly productivity measure inflation
causes a difference of $2.5$ in \textit{m}-quotient between average 40
year olds and average 65 year olds.  With differences among
sub-disciplines also a factor of two or more, independent of age, we
suggest that citation counts and derived measures such as \textit{h}
and \textit{m} should not be used, except for crude evaluations of
scholars' impact.

\subsection*{A measure of research impact independent of historical
  \newline and disciplinary effects}
Here we propose a novel, simple, and effective measure of research
performance, designed to minimize the disciplinary and historical
effects which most negatively affect citation counts and derivative
measures.  In addition to the volume of citations,
the proposed measure employs two more bits of bibliographic information,
readily available by modern scholarly databases: the
number of authors and the number of references in a paper.

Both these measures have been used before, separately.  Adjusting
citation counts for the number of authors seems obvious
\cite{desollaprice79}, and has been available as an option in the ADS
system since 1996 \cite{kurtz05b}.  Adjusting for the number of
references has become a standard technique in evaluating journals,
with Web of Science using Eigenfactor \cite{west} and SCOPUS using SNIP
\cite{moed}.  Similar normalizations of the \textit{h} and other
indices have been proposed in the literature, as discussed above. For example, dividing
the \textit{h}-index by the number of authors in a paper
\cite{batista} and by the average number of citations per article per
discipline \cite{radicchi} both yield promising results for
cross-disciplinary impact comparison.

Thus, we normalize every external (non-self) citation received by a scholar
in two ways: by the number of authors in the cited paper and by the
number of references in the citing article. We speculate that {\it a
simple double normalization, by number of authors and by number of
references in the citing article, has the effect of grounding
productivity index in the authorship and citation practices of a given
field at a given time.}

We define the Total Research Impact, {\it tori}, of a scholar as:

%%%
\begin{equation}
\label{formula:ror}
tori = \displaystyle\sum_{n} \frac{1}{a \cdot r}
\end{equation}
%%% 

where $n$ is the collection of external (non-self) citations accrued
by the researcher, $a$ is the number of authors of the cited paper,
and $r$ is the number of bibliographic references of the citing
paper. One calculates the overall, cumulative output of a scholar by
summing the impact of every external citation accrued in his/her
career. As such, the total research impact of a scholar ({\it tori}) is
simply defined as \textit{the amount of work that others have devoted
to his/her research, measured in research papers}.

The definition of {\it tori} influences the self-citation correction.  The
standard self-citation correction \cite{Wuchty} removes a citation
if any of the authors of the citing paper are the same as the authors
of the cited paper. With the computation of {\it tori}, we only remove a citation {\it if the
author being measured is an author of the citing paper.} \cite{glanzel}

We can also compute the research impact averaged over a scholar's
career, equivalent to the \textit{m}-quotient. For a scholar with a
career span of $y$ years, the Research Impact Quotient, \textit{riq},
is defined as:

%%%
\begin{equation}
\label{formula:ror}
riq = \frac{\sqrt{tori}}{{y}}
\end{equation}
%%% 

\begin{figure*}[hp!]
\begin{center}
\includegraphics[width=1\textwidth]{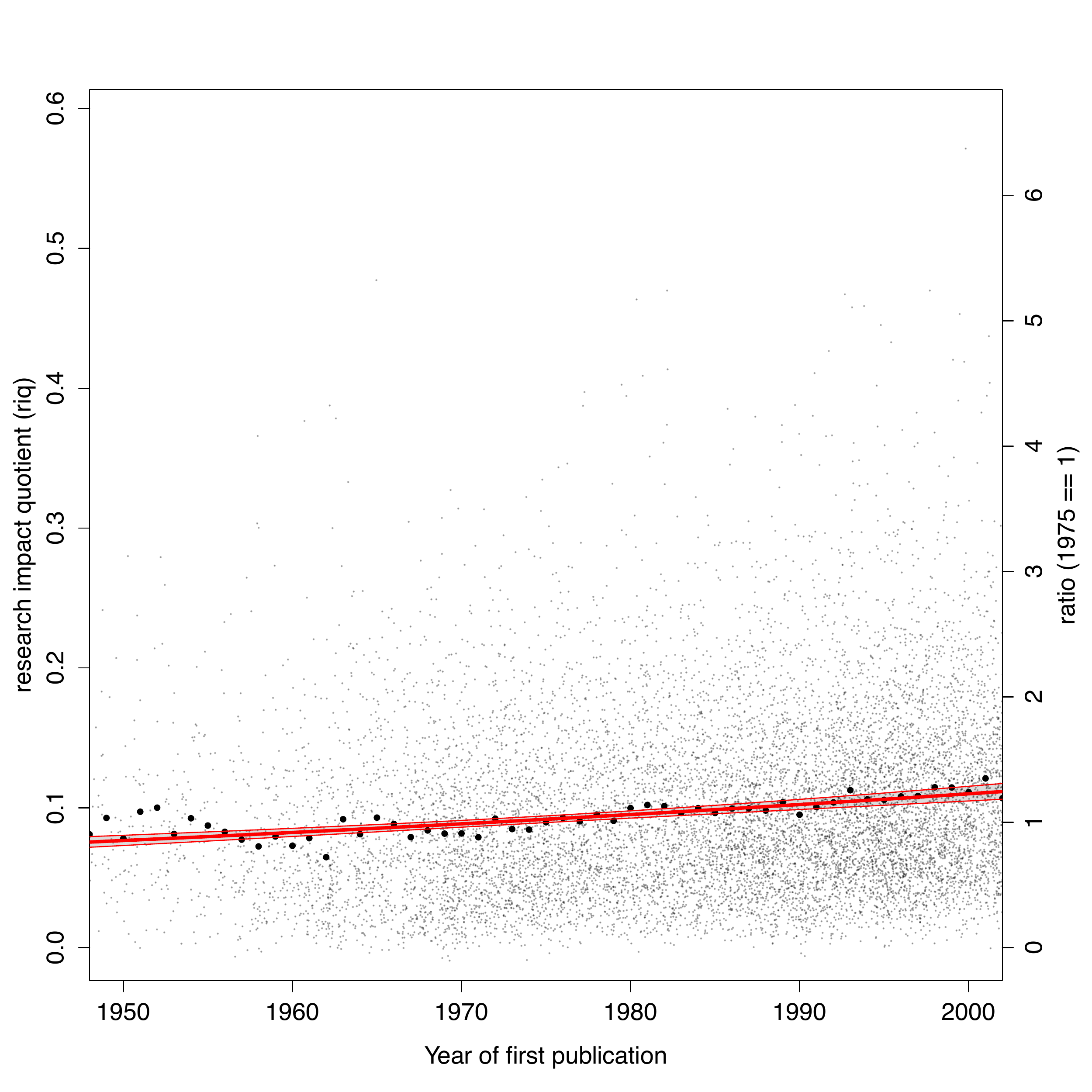}
\caption{\label{fig:riqVSy1} Distribution of astronomers' research
  impact quotients as function
  of beginning of career (defined as the year of first refereed publication).}
\end{center}
\end{figure*}

\begin{figure*}[hp!]
\begin{center}
\includegraphics[width=1\textwidth]{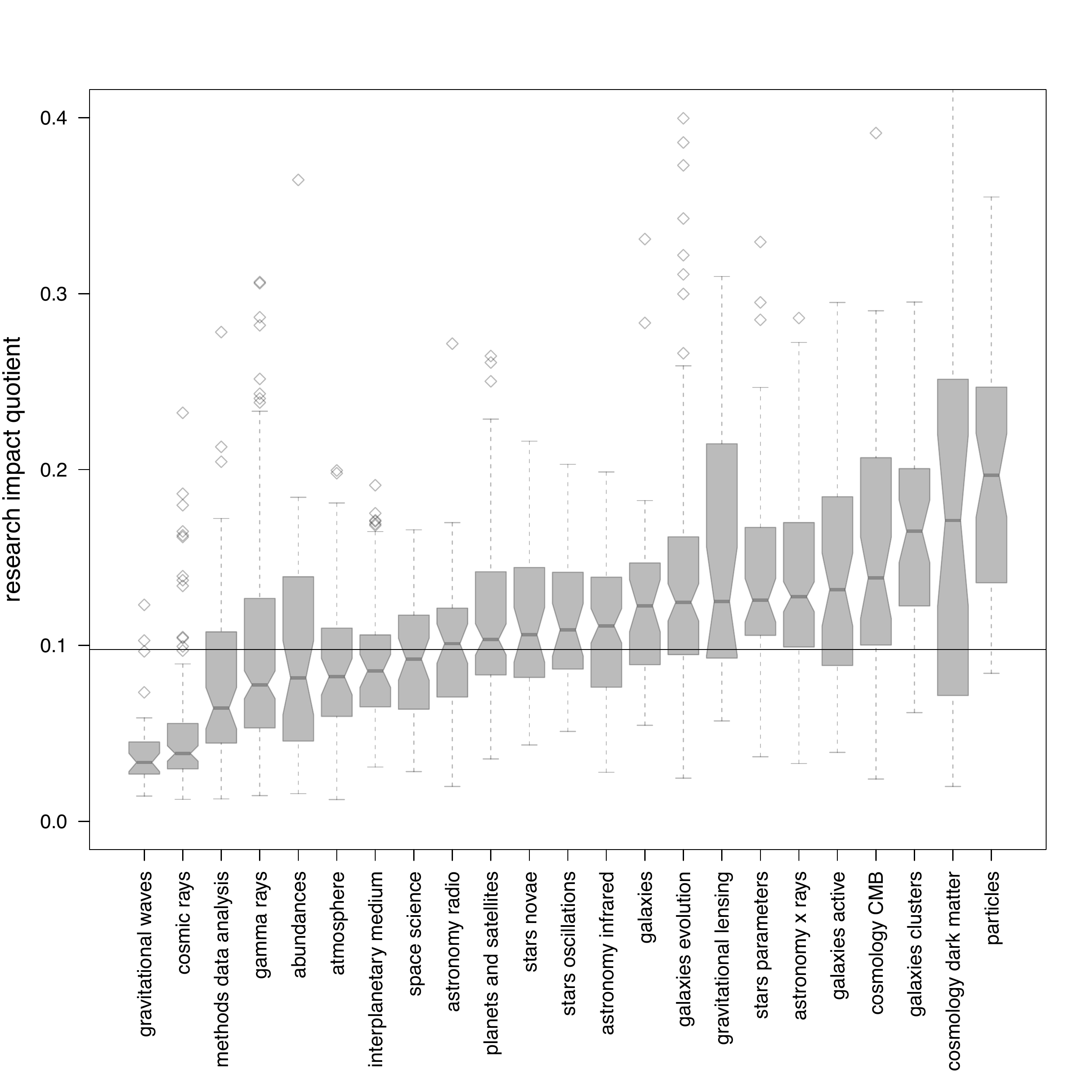}
\caption{\label{fig:riqVSdisc} Distribution of astronomers' research
  impact quotient (riq) as function of field of specialization. Each field includes between 30 and 150 astronomers who began their career in the 1990s. The box-and-whisker plot of each field depicts the median (middle notch), lower and upper quartiles (lower and upper hinges), minimum and maximum values (lower and upper whiskers), and outliers.}
\end{center}
\end{figure*}

We test the performance of this measure on the same corpus discussed
above, finding that the research output quotient performs very well
both over time and across sub-disciplines of astronomy, as shown in
Figures \ref{fig:riqVSy1} and \ref{fig:riqVSdisc}.

Temporal debasement effects are greatly attenuated when computing the
\textit{riq} on this population of scholars. As shown in Figure
\ref{fig:riqVSy1}, astronomers who began their career in the 1950s do
perform, on average, similar to astronomers who started publishing 50
years later (global mean is $x = 0.098$).  An exponential best-fit
regression line (shown as solid line, with a $0.95$ confidence band)
still shows a positive gradient ($b = 0.00659$), but considerably
smaller than that of {\it m}. (A similar analysis on a cohort of 544
astronomy Ph.D. confirmed this result, finding an exponential
regression line with slope $b = 0.00417$).  The large attenuation of
temporal effects obtained with the computation of {\it riq} is not
predominately due to either of the two normalizations: they both
contribute roughly equally.  The temporal slope after removing the
effects of multiple co-authors is $b = 0.015$ and the slope after the
normalization by number of references only is $b = 0.020$.

The disciplinary bias observed for the \textit{m}-quotient, previously
discussed and depicted in Figure \ref{fig:mVSdisc}, are greatly
improved when the \textit{riq} is computed, as shown in Figure
\ref{fig:riqVSdisc}. While astronomers working in certain disciplines
do perform below (i.e., ``gravitational waves'' and ``cosmic rays'')
or above average (i.e., ``stars parameters'', ``galaxies clusters'', and
``particles''), the lower and upper {\it riq} quartile band
measured for the majority of
fields analyzed (18 out of 23) tends to fall within the global mean
\textit{riq} ($x = 0.098$, shown as a dashed line).

\subsection*{Direct comparison of m-quotient and riq}
\begin{figure*}[hp!]
\begin{center}
\includegraphics[width=0.9\textwidth]{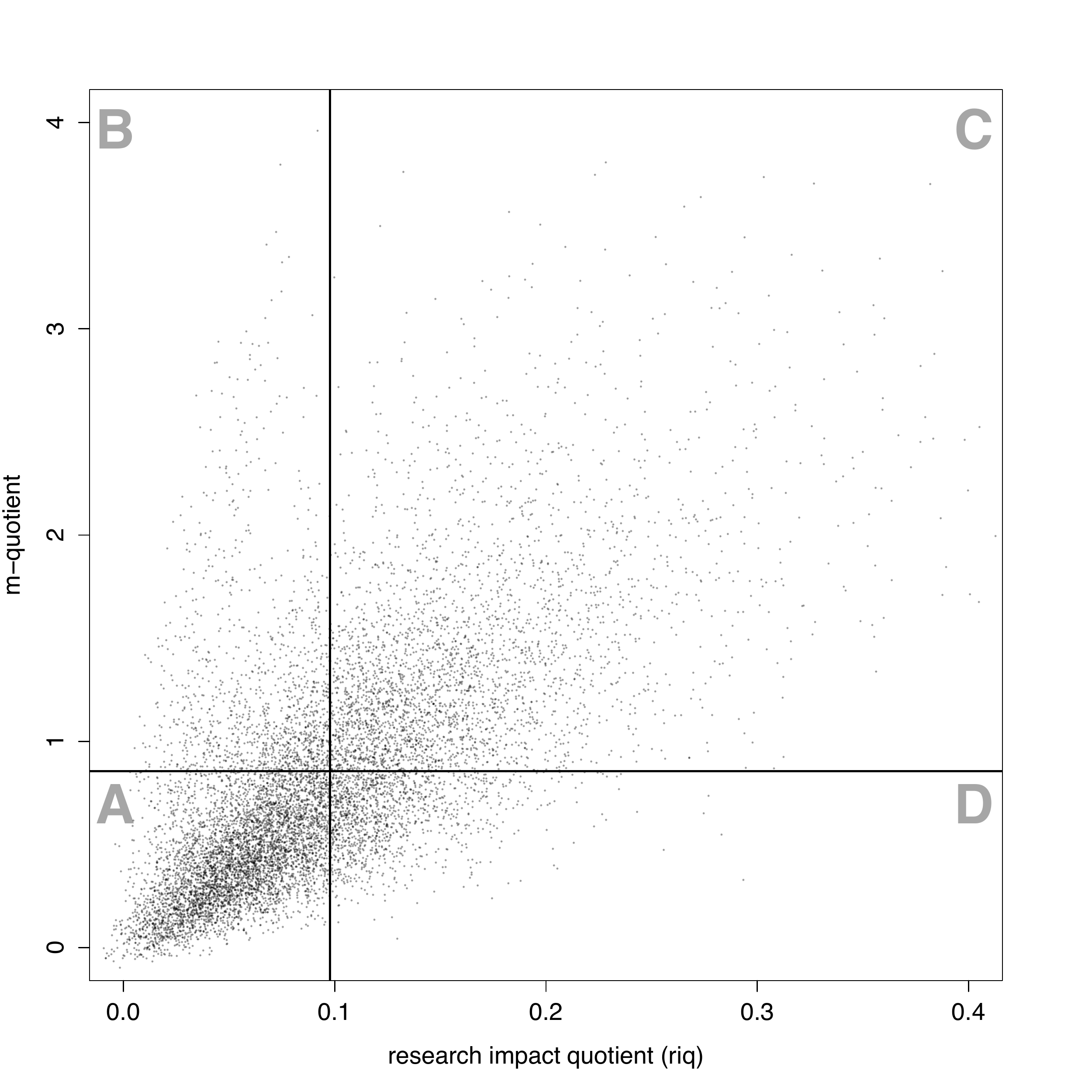}
\caption{\label{fig:mVSriq} Scatterplot of {\it m} vs. {\it riq}. Horizontal and vertical lines depict the global mean {\it m} and {\it riq}, respectively.}
\end{center}
\end{figure*}

To better illustrate the differences between the two measures
discussed here, in Figure \ref{fig:mVSriq}, we present a scatterplot
of {\it m} versus {\it riq} for each astronomer in the corpus. The
solid horizontal and vertical lines indicate the global mean for {\it
m} and {\it riq}, respectively. By and large, {\it m} and {\it riq}
are positively correlated, but with a substantial scatter on both
sides of the main correlation trend. Moreover, an anomaly of the
scatter plot is the presence of a collection of points in the upper
left B quadrant: they form a branch which does not follow the
main overall trend. In quadrant B, we identify $1221$ astronomers who have {\it m} above the global mean
($m > 0.855$), but {\it riq} below the global mean ($riq <
0.0977$). In the same way, we isolate astronomers in the lower
right quadrant indicated as ``D'', who have above mean {\it riq } and
below mean {\it m}. Although the scatter in quadrant D
is much less prominent, this group of $1300$ astronomers has {\it m}
below the mean and {\it riq} above the mean. Astronomers in the upper
left (B) and lower right (D) quadrants are interesting to explore more
in detail as they are weighed very differently by the two productivity
measures. Some descriptive statistics about these groups, and the
overall population, are presented in Table \ref{tab:desc}. 

\begin{table*}[h!]
\caption{Descriptive statistics regarding astronomers with {\it m}
  above the mean and {\it riq} below the mean (quadrant B),
  astronomers with {\it riq} above the mean and {\it m} below the mean
  (quadrant D), and the overall
population (all quadrants)\vspace{0.8em}}
\label{tab:desc}
\centering
\begin{tabular*}{1\textwidth}
{@{\extracolsep{\fill}}lrrr}
%{1\textwidth}{@{\extracolsep{\fill}}|l|r|r|}
\hline
&quadrant \textbf{B}& quadrant \textbf{D}  & \textbf{all} quadrants \\
\hline\hline
size&$1,221$\phantom{.00}&$1,300$\phantom{.00}&$11,036$\phantom{.00}\\
number of publications&$59.79$&$71.86$&$64.02$\\
number of first authored publications&$5.94$&$17.35$&$12.57$\\
career start, year&$1,993.1$\phantom{0}&$1,977.1$\phantom{0}&$1,983.5$\phantom{0}\\
career length, years&$17.6$\phantom{0}&$31.4$\phantom{0}&$25.9$\phantom{0}\\
citations accrued, total&$2,078.5$\phantom{0}&$1,666.4$\phantom{0}&$1,931.9$\phantom{0}\\
citations accrued, normalized&$116.8$\phantom{0}&$654.4$\phantom{0}&$429.8$\phantom{0}\\
{\it h}-index&$21.73$&$19.78$&$19.12$\\
{\it tori}&$1.81$&$19.66$&$9.55$\\
\hline
\end{tabular*}
\end{table*}

Table \ref{tab:desc}
shows that astronomers in quadrant B and D publish differently. In
quadrant D, we find astronomers who publish profusely (well above the global
mean), both in general and as first authors. Astronomers in quadrant B
not only publish below the mean, but also publish only a very small
fraction of them as first authored works (1 in 12, as opposed to 1 in
4 for quadrant B and 1 in 5 for the overall population). Looking at
the careers of astronomers in the two groups, we find that those in
quadrant B tend to be younger scholars with shorter career time spans,
than those in quadrant D. The citations accrued by astronomers in
these two groups also follow different dynamics, with astronomers in
quadrant B receiving a large volume of citations, although citation impact drops
substantially below the mean if accrued citations are
normalized by the number of authors in a paper. Quadrant D follows a
perfectly inverse pattern: fewer overall citations, but more
normalized citations. Finally, a look at the research productivity
indices for the two groups reveals that quadrant B astronomers have on
average higher {\it h} and considerably lower {\it tori} than the
global mean (and vice versa for quadrant D). These effects, especially
those relative to citation and publication, are indicative of the
different archetypes of astronomers that are found in the two
sections: quadrant B scholars are part of highly cited, large
collaborations; quadrant D scholars are part of highly cited, yet
smaller collaboration groups. A detailed examination of the careers
of the individuals who are the most extreme outliers confirms this
analysis.

\section*{Discussion}
The discussed measures --- {\it tori} and {\it riq} --- eliminate the most important systematic factors
affecting the use of citations to measure the performance of
individuals: the number of authors of each paper, the number of
references in each paper, and the age of the individual.  In addition,
they remove the self citation bias.  The shot noise problem is not
directly addressed, however it is essentially eliminated by the number
of authors correction \cite{Spruit}.  The problem of curation was
addressed in this study by careful selection of non-ambiguous names; it is
being addressed more generally by initiatives such as
ORCID \cite{orcid}.

Both {\it tori} and {\it riq} are designed to measure individuals; aggregations
of individuals such as countries, universities, and departments, can be
characterized by simple summary statistics, such as the number of
scientists and their mean {\it riq}. An extension of {\it tori} to measure
journals would be straight forward: it would consist of the simple
removal of the normalization by the number of authors.  The result would
be similar to SNIP \cite{moed}; we suggest that SNIP and Eigenfactor \cite{west} continue to be used
for the purpose of measuring journals.

%Despite the advantages of tori and {\it riq} over existing indices,

While {\it tori} and {\it riq} remove the largest systematic problems with
citation counts, they are citation-derived measures and, as such, 
they necessarily suffer from two systematic problems of citation
counts which
do not lend themselves to programmatic solutions. First, it is not in
general possible to tell the differing contributions of various
co-authors to a paper ({\it tori} assumes all authors contribute equally ---
a technique obviously more correct in the aggregate than for any individual paper).
Modifications, such as giving extra weight to the first (or last) author are necessarily ad hoc and discipline-dependent stratagems. The second fundamental problem with the use of citations for the evaluation of
individuals is that citations chiefly measure usefulness
\cite{Nicolaisen}, \textit{not} importance; {\it tori} is no exception.
While usefulness can be correlated with importance, these are clearly
different concepts; oftentimes, importance is what is actually desired.

%tori and {\it riq} already remove the largest systematic problems
%with citation counts, further corrections are not likely to be
%significant when compared with the other factors taken into
%consideration in any read evaluation of individual performance.

Measuring the research performance of scholars
is a delicate and controversial procedure. That a scientists's career output
cannot be condensed in a bare number is beyond discussion. 
Yet, providing an accurate and concise quantitative
indication of a scholar's individual research output is important, and  
oftentimes necessary. For a faculty member at the early stages of her
career, for example, a quantitative indication of her scientific
productivity can be the factor determining whether she will be
promoted to tenure or whether she will be awarded a research grant. In
some other contexts, it may be crucial for funding bodies to know the aggregated research
output of individuals working in academic institutions and scientific
organizations as this can affect the course of science policy
decisions. 

But, most importantly, measuring research output with
accuracy is important chiefly because \textit{scholars themselves are
interested in knowing their own research performance and impact}.
The measures discussed here --- {\it tori} and {\it riq} --- while not easily computed by an individual, are
easily derived from information already available in virtually all bibliographic
repositories, such as Web of
Science (\url{http://wokinfo.com/}),
Scopus (\url{http://www.scopus.com}), SciFinder (\url{http://www.cas.org/products/scifindr/}),
and ACM-DL (\url{http://dl.acm.org/}).
The Astrophysics Data System (ADS) (\url{http://adsabs.org/}) has already implemented them and
they are currently available to the entire
community of astronomers and astrophysicists. We suggest that these
measures become part other academic databases as well, to allow a
fair measurement of researchers' output across time and disciplines. 

\section*{Acknowledgments}
We thank Alberto Accomazzi and the SAO/NASA Astrophysics Data System team
  at the Harvard-Smithsonian Center for Astrophysics for providing
  access to the bibliographic data. The research content of this
  article was discussed with Alyssa Goodman (Harvard-Smithsonian
  Center for Astrophysics) and Matteo Cantiello (Kavli Institute for
  Theoretical Physics, University of California, Santa Barbara).

\bibliography{scibib}
\end{document}